\documentstyle[12pt]{article}
\def\be {\begin{equation}}
\def\ee {\end{equation}}
\def\ba {\begin{eqnarray}}
\def\ea {\end{eqnarray}}
\newcommand{\bq}{\begin{eqnarray}}
\newcommand{\eq}{\end{eqnarray}}

\def\bi {\begin{itemize}}
\def\ei {\end{itemize}}
\begin{document}
\def\bea{\begin{eqnarray}}
\def\eea{\end{eqnarray}}
\title{\bf {Holographic Chaplygin DGP cosmologies}}
 \author{M.R. Setare  \footnote{E-mail: rezakord@ipm.ir}
  \\ {Department of Science,  Payame Noor University. Bijar, Iran}}
\date{\small{}}
\maketitle
\begin{abstract}
In the present paper, we present an extra dimensions inspired model
that is built on the DGP brane-world scenario, then we take the dark
energy component on the brane to be a Chaplygin gas. After that we
consider a holographic model of Chaplygin gas in the framework of
DGP cosmology. We show that the holographic Chaplygin gas can mimic
a phantom fluid and cross the phantom divide in a DGP brane-world
setup.
 \end{abstract}

\newpage
\section{Introduction}
 The type Ia supernova observations
suggests that the universe is dominated by dark energy (DE) with
negative pressure which provides the dynamical mechanism of the
accelerating expansion of the universe \cite{{per},{gar},{ries}}.
The strength of this acceleration is presently matter of debate,
mainly because it depends on the theoretical model implied when
interpreting the
data.\\
An approach to the problem of DE arises from the holographic
principle that states that the number of degrees of freedom related
directly to entropy scales with the enclosing area of the system. It
was shown by 'tHooft and Susskind \cite{hologram} that effective
local quantum field theories greatly overcount degrees of freedom
because the entropy scales extensively for an effective quantum
field theory in a box of size $L$ with UV cut-off $ \Lambda$. As
pointed out by \cite{myung}, attempting to solve this problem, Cohen
{\it et al} showed \cite{cohen} that in quantum field theory, short
distance cut-off $\Lambda$ is related to long distance cut-off $L$
due to the limit set by forming a black hole. In other words the
total energy of the system with size $L$ should not exceed the mass
of the same size black hole, i.e. $L^3 \rho_{\Lambda}\leq LM_p^2$
where $\rho_{\Lambda}$ is the quantum zero-point energy density
caused by UV cut-off $\Lambda$ and $M_P$ denotes the Planck mass (
$M_p^2=1/{8\pi G})$. The largest $L$ is required to saturate this
inequality. Then its holographic energy density is given by
$\rho_{\Lambda}= 3c^2M_p^2/ L^2$ in which $c$ is a free
dimensionless parameter and coefficient 3 is for convenience. As an
application of the holographic principle in cosmology,
 it was studied by \cite{KSM} that the consequence of excluding those degrees of freedom of the system
 which will never be observed by the effective field
 theory gives rise to IR cut-off $L$ at the
 future event horizon. Thus in a universe dominated by DE, the
 future event horizon will tend to a constant of the order $H^{-1}_0$, i.e. the present
 Hubble radius.
 On the basis of the cosmological state of the holographic principle, proposed by Fischler and
Susskind \cite{fischler}, a holographic model of dark Energy (HDE)
has been proposed and studied widely in the
 literature \cite{miao,HDE} \footnote{ A very recent development
 is the idea of bulk holographic dark energy.
In this proposal, holographic dark energy is accommodated in the
framework of braneworld cosmology \cite{Saridakis:2007cy}.} In HDE,
in order to determine the proper and well-behaved system's IR
cut-off, there are some difficulties that must be studied carefully
to get results adapted with experiments that claim our universe has
accelerated expansion. For instance, in the model proposed by
\cite{miao}, it is discussed that considering  the particle horizon,
as the IR cut-off, the HDE density reads
 \be
  \rho_{\Lambda}\propto a^{-2(1+\frac{1}{c})},
\ee
 that implies $w>-1/3$ which does not lead to an accelerated
universe. Also it is shown in \cite{easther} that for the case of
closed
universe, it violates the holographic bound.\\
The problem of taking apparent horizon (Hubble horizon) - the
outermost surface defined by the null rays which instantaneously are
not expanding, $R_A=1/H$ - as the IR cut-off in the flat universe
was discussed by Hsu \cite{Hsu}. According to Hsu's argument,
employing the Friedmann equation $\rho=3M^2_PH^2$ where $\rho$ is
the total energy density and taking $L=H^{-1}$ we will find
$\rho_m=3(1-c^2)M^2_PH^2$. Thus either $\rho_m$ or $\rho_{\Lambda}$
behave as $H^2$. So the DE results as pressureless, since
$\rho_{\Lambda}$ scales like matter energy density $\rho_m$ with the
scale factor $a$ as $a^{-3}$. Also, taking the apparent horizon as
the IR cut-off may result in a constant parameter of state $w$,
which is in contradiction with recent observations implying variable
$w$ \cite{varw}. On the other hand taking the event horizon, as
 the IR cut-off, gives results compatible with observations for a flat
 universe.\\
 It is fair to claim that the simplicity and reasonable nature of HDE
 provide a
 more reliable framework  for investigating the problem of DE compared with other models
proposed in the literature\cite{cosmo,quint,phant}. For instance the
coincidence or "why now?" problem is easily solved in some models of
HDE based on this fundamental assumption that matter and holographic
dark energy do not conserve separately, but the matter energy
density
decays into the holographic energy density \cite{interac}.\\
 In a
very interesting paper Kamenshchik, Moschella, and Pasquier
\cite{kmp}have studied a homogeneous model based on a single fluid
obeying the Chaplygin gas equation of state \be \label{chp}
P=\frac{-A}{\rho} \ee where $P$ and $\rho$ are respectively pressure
and energy density in comoving reference frame, with $\rho> 0$; $A$
is a positive constant. This equation of state has raised  a certain
interest \cite{jac} because of its many interesting and, in some
sense, intriguingly unique features. Some possible motivations for
this model from the field theory points of view are investigated in
\cite{a}. The Chaplygin gas emerges as an effective fluid associated
with d-branes \cite{b} and can also be obtained from the Born-Infeld
action \cite{c}.\\
An alternative way of explaining the observed acceleration of the
late universe is to modify gravity at large scales. In the present
paper, we present an extra dimensions inspired model that is built
on the DGP braneworld scenario \cite{2}, in this framework,
existence of a higher dimensional embedding space allows for the
existence of bulk or brane matter which can certainly influence the
cosmological evolution on the brane. Here we take the dark energy
component on the brane to be a Chaplygin gas.  The DGP model has a
large scale/low energy effect of causing the expansion rate of the
universe to accelerate. We assume that the relation $\rho_{\Lambda}=
3c^2M_p^2/ L^2$ still holds in the DGP model, then we suggest  a
correspondence between the holographic dark energy scenario and the
Chaplygin gas dark energy model in the framework of DGP cosmology.
According to the result of \cite{mar} a Chaplygin gas can mimic a
phantom fluid and cross the phantom divide in a DGP brane-world
setup. Our calculation show, taking $\Omega_{\Lambda}=0.73$ for the
present time, it is possible to have $w_{\rm \Lambda}$ crossing
$-1$. This implies that one can generate phantom-like equation of
state from the holographic dark energy model in flat universe in the
DGP cosmology framework.
 We show this holographic description of the
Chaplygin gas dark energy in FRW universe  and reconstruct the
potential and the dynamics of the scalar field which describe the
Chaplygin cosmology. We show in order that the holographic dark
energy model, which is inspired by quantum gravity, to be consistent
with the Chaplygin gas model of dark energy in the DGP framework,
one must be careful to satisfy the corresponding constraints which
given by equations. (30), (31), i.e, the scalar field which is the
origin of Chaplygin gas has to follow potential (30) and equation
(31), in order for the two models to be compatible and thus
efficiently unified.

\section{ Holographic dark energy and Chaplygin gas in DGP braneworld }
We consider a DGP- braneworld model where the dark energy component
on the brane is given by a Chaplygin gas, and with an extra CDM
component \cite{mar}. In the DGP model, it is supposed that a
3-dimensional brane is embedded in 5-dimensional spacetime. This
model predicts that 4-dimensional Einstein gravity is a
short-distance phenomenon with deviations showing up at large
distances. The DGP model includes a length scale below which the
potential has usual Newtonian form and above which the gravity
becomes 5-dimensional. The cross over scale between the
4-dimensional and 5-dimensional gravity is  given by
\begin{equation}\label{5}
r_{c}=\frac{M_{P}^{2}}{2M_{5}^{2}}
\end{equation}
where $M_{5}$ is the fundamental scale of gravity in five dimension.
 For the spatially flat
Robertson-Walker universe
\begin{equation}\label{met}
ds^{2}=-dt^{2}+a(t)^{2}(dr^{2}+r^{2}d\Omega^{2}).
\end{equation}
the first Friedmann equation is given by
\begin{equation}\label{3}
3H^2=\rho_{m}+\rho_{eff}.
\end{equation}
here we take $M_{P}^{2}=1$, and $\rho_{m}$ is the energy density of
CDM and effective energy density is given by
\begin{equation}\label{4}
\rho_{eff}=\rho-\frac{3H}{r_{c}}.
\end{equation}
 where $\rho$ is the energy density of Chaplygin gas.
Inserting the equation of state (\ref{chp}) into the relativistic
energy conservation equation, leads to a density evolving as \be
\label{enerd}\rho=\sqrt{A+\frac{B}{a^{6}}} \ee where $B$ is an
integration constant.\\
Now following \cite{bar} we assume that the origin of the dark
energy is a scalar field $\phi$, so \be \label{roph1}
\rho_{\phi}=\frac{1}{2}\dot{\phi}^{2}+V(\phi)=\sqrt{A+\frac{B}{a^{6}}}
\ee \be \label{roph2}
P_{\phi}=\frac{1}{2}\dot{\phi}^{2}-V(\phi)=\frac{-A}{\sqrt{A+\frac{B}{a^{6}}}}
\ee Then, one can easily derive the scalar potential and kinetic
energy term as \be
\label{vph}V(\phi)=\frac{2a^6(A+\frac{B}{a^{6}})-B}{2a^6\sqrt{A+\frac{B}{a^{6}}}}
\ee \be \label{phid}
\dot{\phi}^{2}=\frac{B}{a^6\sqrt{A+\frac{B}{a^{6}}}} \ee
 Now we
suggest a correspondence between the holographic dark energy
scenario and the  Chaplygin gas in DGP braneworld. The holographic
dark energy scenario reveals the dynamical nature of the vacuum
energy. When taking the holographic principle into account, the
vacuum energy density will evolve dynamically. On the other hand,
the scalar field dark energy models are often viewed as effective
description of the underlying theory of dark energy. We are now
interested in that if we assume the holographic vacuum energy
scenario as the underlying theory of dark energy, how the scalar
field model can be used to effectively describe it.\\
 Our choice for
holographic dark energy density is
 \be
\rho_{\Lambda}=\frac{3c^2}{R_{h}^2} \label{holo}\ee where $c$ is a
constant, and $R_h$ is the future event horizon given by \be
  R_h= a\int_t^\infty \frac{dt}{a}=a\int_a^\infty\frac{da}{Ha^2}
 \ee
 The critical energy density, $\rho_{cr}$, is given by following relation
\begin{eqnarray} \label{ro}
\rho_{cr}=3H^2
\end{eqnarray}
Now we define the dimensionless dark energy as \be
\Omega_{\Lambda}=\frac{\rho_{\Lambda}}{\rho_{cr}}=\frac{c^2}{R_{h}^2H^2}\label{omega}
\ee Using definition $\Omega_\Lambda$ and relation (\ref{ro}),
$\dot{R_{h}}$ gets: \be \label{ldot} \dot{R_{h}} =
R_{h}H-1=\frac{c}{\sqrt{\Omega_\Lambda}}-1,
\end{equation}
By considering  the definition of holographic energy density
$\rho_{\rm \Lambda}$, and using Eqs.( \ref{omega}), (\ref{ldot}) one
can find:
\begin{equation}\label{roeq}
\dot{\rho_{\Lambda}}=-2H(1-\frac{\sqrt{\Omega_\Lambda}}{c})\rho_{\Lambda}
\end{equation}
Substitute this relation into following equation
\begin{eqnarray}
\label{2eq1}&& \dot{\rho}_{\rm \Lambda}+3H(1+w_{\rm
\Lambda})\rho_{\rm \Lambda} =0,
\end{eqnarray}
 we obtain
\begin{equation}\label{stateq}
w_{\rm \Lambda}=-(\frac{1}{3}+\frac{2\sqrt{\Omega_{\rm
\Lambda}}}{3c}) .
\end{equation}
A direct fit of the present available SNe Ia data with this
holographic model indicates that the best fit result is $c=0.21$
\cite{HG}. Recently, by calculating the average equation of state of
the dark energy and the angular scale of the acoustic oscillation
from the BOOMERANG and WMAP data on the CMB to constrain the
holographic dark energy model, the authors show that the reasonable
result is $c\sim 0.7$ \cite{cmb1}. In the other hand, in the study
of the constraints on the dark energy from the holographic
connection to the small $l$ CMB suppression, an opposite result is
derived, i.e. it implies the best fit result is $c=2.1$ \cite{cmb3}.
Thus according to these studies $0.21\leq c\leq 2.1$.  Taking
$\Omega_{\Lambda}=0.73$ for the present time, in the case of
$c=0.21$, we obtain $w_{\rm \Lambda}=-3.04$, in the other hand for
$c=2.1$, one can obtain, $w_{\rm \Lambda}=-0.6$. Using
Eq.(\ref{stateq}), one can see that by considering $c\leq
\sqrt{\Omega_{\Lambda}}$ we obtain $w_{\rm \Lambda}\leq -1$.
Therefore taking $\Omega_{\Lambda}=0.73$ for the present time, it is
possible to have $w_{\rm \Lambda}$ crossing $-1$. Also the authors
of \cite{mar} have recently showed that a Chaplygin gas can mimic a
phantom fluid and cross the phantom divide in a DGP brane-world
setup. \\If we establish the correspondence between the holographic
dark energy and effective energy density in DGP braneworld
\begin{equation}\label{18}
\frac{3c^2}{R_{h}^{2}}=\sqrt{A+\frac{B}{a^{6}}}-\frac{3H}{r_{c}} .
\end{equation}
then using Eqs.(\ref{omega}),(\ref{18}) we have \be
\label{beq}B=a^6[9(\Omega_{\rm \Lambda}H^{2}+\frac{H}{r_{c}})^{2}-A]
\ee Now we define the effective equation of state parameter
$w_{eff}$ as \cite{mar} \be
\label{21}1+w_{eff}=\frac{-\dot{\rho}_{eff}}{3H\rho_{eff}}\ee Using
Eqs. (\ref{chp}), (\ref{4}) and following continuity equation for
Chaplygin gas \be \dot{\rho}+3H(\rho+P)=0 \label{coneq}\ee one can
obtain following equation
 \be \label{23}\dot{\rho}_{eff}=\frac{9HH_{0}^{2}(1+Z)^{3}}{H+H_0\sqrt{\Omega_{rc}}}[\Omega_{m}
 H_0
 \sqrt{\Omega_{rc}}-H(1-\frac{A}{\rho_{0}^{2}})\Omega(1+Z)^{3}[\frac{A}{\rho_{0}^{2}}+(1-\frac{A}{\rho_{0}^{2}})
 (1+Z)^{6}]^{-1/2}]
 \ee
 where the subscript $0$
corresponds to the current value of a given quantity, $Z = (1/a) -
1$ is the redshift of the universe, and
\begin{equation}\label{24}
\Omega_{m}=\frac{\rho_{m0}}{3H_{0}^{2}}, \hspace{1cm}
\Omega=\frac{\rho_{0}}{3H_{0}^{2}},
\hspace{1cm}\Omega_{rc}=\frac{1}{4r_{c}^{2}H_{0}^{2}}.
\end{equation}
Using the correspondence between the holographic dark energy and
effective energy density in DGP braneworld we claim that
$w_{eff}=w_{\Lambda}$, as we have mentioned above, if
$c\leq\sqrt{\Omega_{\Lambda}}$, then $w_{eff}$ crosses
$-1$.\\
 Using Eqs.(\ref{stateq}), (\ref{21}), and
(\ref{23}) we can obtain
\begin{equation}\label{25}
A=\rho_{0}\frac{(d\pm\sqrt{d^2+4e})}{2}
\end{equation}
where
\begin{equation}\label{26}
d=2+e(\frac{1}{(1+Z)^{6}}-1)
\end{equation}
and \begin{equation}\label{27}
e=\frac{\Omega_{m}H_0\sqrt{\Omega_{rc}}}{H\Omega}-\frac{2\Omega_{\Lambda}\frac{\sqrt{\Omega_{\Lambda}}}{c}
H+H_0\sqrt{\Omega_{rc}}}{3\Omega(1+Z)^{3}H_{0}^{2}}
\end{equation}
Substituting $A$ into Eq.(\ref{beq}) we obtain following relation
for $B$
 \be
\label{beq1}B=a^6[9(\Omega_{\rm
\Lambda}H^{2}+\frac{H}{r_{c}})^{2}-\rho_{0}\frac{(d\pm\sqrt{d^2+4e})}{2}]
\ee
  Now we can
rewritten the scalar potential and kinetic energy term as following
\bq \label{vphi1} V(\phi)=3H^2\Omega_{\Lambda}-\frac{9(\Omega_{\rm
\Lambda}H^{2}+\frac{H}{r_{c}})^{2}-\frac{\rho_{0}(d\pm\sqrt{d^2+4e})}{2}}{6H^2\Omega_{\Lambda}}
\eq \be \label{phi2} \dot{\phi}=[\frac{9(\Omega_{\rm
\Lambda}H^{2}+\frac{H}{r_{c}})^{2}-\frac{\rho_{0}(d\pm\sqrt{d^2+4e})}{2}}{3H^2\Omega_{\Lambda}}]^{1/2}
\ee Considering $x(\equiv lna)$, we have \be \label{phid}
\dot{\phi}=\phi' H \ee Then derivative of scalar field $\phi$ with
respect to $x(\equiv lna)$ is as \be \label{phi3}
\phi'=[\frac{9(\Omega_{\rm
\Lambda}H^{2}+\frac{H}{r_{c}})^{2}-\frac{\rho_{0}(d\pm\sqrt{d^2+4e})}{2}}{3H^4\Omega_{\Lambda}}]^{1/2}
\ee Consequently, we can easily obtain the evolutionary form of the
field \be \label{phi4}\phi(a)-\phi(a_0)=\int_{0}^{\ln a}
[\frac{9(\Omega_{\rm
\Lambda}H^{2}+\frac{H}{r_{c}})^{2}-\frac{\rho_{0}(d\pm\sqrt{d^2+4e})}{2}}{3H^4\Omega_{\Lambda}}]^{1/2}
dx \ee where $a_0$ is the present time value of the scale factor.
\section{Conclusions}
A well-studied model of modified gravity is the DGP braneworld model
\cite{2} in which our $4-$dimensional world is a FRW brane embeded
in a $5-$dimensional Minkowski bulk. In the other hand within the
different candidates to play the role of the dark energy, the
Chaplygin gas, has emerged as a possible unification of dark matter
and dark energy, since its cosmological evolution is similar to an
initial dust like matter and a cosmological constant for late times.
Inspired by the fact that the Chaplygin gas possesses a negative
pressure, people \cite{mas} have undertaken the simple task of
studying a FRW cosmology of a universe filled with this type of
fluid.\\
In this paper we have considered a holographic model of Chaplygin
gas in the framework of DGP cosmology. We have shown that the
holographic dark energy can be described  by the scalar field in a
certain way. Then a correspondence between the holographic dark
energy and Chaplygin gas model of dark energy in the framework of
DGP cosmology has been established, and the potential of the
holographic scalar field and the dynamics of the field have been
reconstructed. We have shown if $c\leq\sqrt{\Omega_{\Lambda}}$, the
holographic dark energy model also will behave like a phantom model
of dark energy the amazing feature of which is that the equation of
state of dark energy component $w_{\rm \Lambda}$ crosses $-1$.
Hence, we see, the determining of the value of $c$ is a key point to
the feature of the holographic dark energy and the ultimate fate of
the universe as well.

\end{document}